\documentclass[letterpaper, 12pt]{article}[2021/05/19]
\usepackage[english]{babel}
\usepackage{amsfonts,amsmath,amssymb,amsthm,latexsym,amscd,mathrsfs}
\usepackage{ifthen,cite}
\usepackage[bookmarksnumbered=true]{hyperref}
\hypersetup{pdfpagetransition={Split}}

\newcommand{\evenhead}{Author \ name}
\newcommand{\oddhead}{Article \ name}
\newcommand{\theArticleName}{Article \ name}

\newcommand{\FirstPageHeading}[1]{\thispagestyle{empty}%
\noindent\raisebox{0pt}[0pt][0pt]{\makebox[\textwidth]{\protect\footnotesize \sf }}\par}

\newcommand{\ArticleName}[1]{\renewcommand{\theArticleName}{#1}\vspace{-2mm}\par\noindent {\LARGE\bf  #1\par}}
\newcommand{\Author}[1]{\vspace{5mm}\par\noindent {\Large  #1\par} \par\vspace{2mm}\par}
\newcommand{\Address}[1]{\vspace{2mm}\par\noindent {\it #1} \par}
\newcommand{\Email}[1]{\ifthenelse{\equal{#1}{}}{}{\par\noindent {\rm E-mail: }{\it  #1} \par}}
\newcommand{\URLaddress}[1]{\ifthenelse{\equal{#1}{}}{}{\par\noindent {\rm URL: }{\tt  #1} \par}}
\newcommand{\EmailD}[1]{\ifthenelse{\equal{#1}{}}{}{\par\noindent {$\phantom{\dag}$~\rm E-mail: }{\it  #1} \par}}
\newcommand{\URLaddressD}[1]{\ifthenelse{\equal{#1}{}}{}{\par\noindent {$\phantom{\dag}$~\rm URL: }{\tt  #1} \par}}

\newcommand{\Abstract}[1]{\vspace{6mm}\par\noindent\hspace*{10mm}
\parbox{140mm}{\small {\bf Abstract.} #1}\par}
\newcommand{\Keywords}[1]{\vspace{3mm}\par\noindent\hspace*{10mm}
\parbox{140mm}{\small {\bf Key words:} \rm #1}\par}
\newcommand{\Classification}[1]{\vspace{3mm}\par\noindent\hspace*{10mm}
\parbox{140mm}{\small {\it 2010 Mathematics Subject Classification:} \rm #1}\vspace{3mm}\par}
\newcommand{\ShortArticleName}[1]{\renewcommand{\oddhead}{#1}}
\newcommand{\AuthorNameForHeading}[1]{\renewcommand{\evenhead}{#1}}

\long\def\@makecaption#1#2{
  \sbox\@tempboxa{\small \textbf{#1.}\ \ #2}%
  \ifdim \wd\@tempboxa >\hsize
    {\small \textbf{#1.}\ \ #2}\par \else
    \global \@minipagefalse
    \hb@xt@\hsize{\hfil\box\@tempboxa\hfil}%
  \fi \vskip\belowcaptionskip}

\def\numberwithin#1#2{\@ifundefined{c@#1}{\@nocounterr{#1}}{%
  \@ifundefined{c@#2}{\@nocnterr{#2}}{%
  \@addtoreset{#1}{#2}%
  \toks@\@xp\@xp\@xp{\csname the#1\endcsname}%
  \@xp\xdef\csname the#1\endcsname
    {\@xp\@nx\csname the#2\endcsname.\the\toks@}}}}
\def\E^#1{{\buildrel #1 \over\vee}}

{\theoremstyle{definition}

}

\usepackage{tikz}

\begin{document}

\FirstPageHeading{V.I. Gerasimenko}

\ShortArticleName{Kinetic equations and hierarchies}

\AuthorNameForHeading{V.I. Gerasimenko}

\ArticleName{\textcolor{blue!50!black}{Kinetic equations and hierarchies of evolution\\ equations of quantum systems}}

\Author{V.I. Gerasimenko$^\ast$\footnote{E-mail: \emph{gerasym@imath.kiev.ua}}}

\Address{$^\ast$Institute of Mathematics of the NAS of Ukraine,\\
    3, Tereshchenkivs'ka Str.,\\
    01601, Kyiv-4, Ukraine}

\bigskip

\Abstract{The article provides an overview of some advances in the mathematical understanding
of the nature of the kinetic equations of quantum systems of many particles. The fundamental
equations of modern mathematical physics are studied, in particular, the hierarchies of
evolution equations of quantum systems and their asymptotic behavior described by kinetic
nonlinear equations.}

\vspace{2pc}
\Classification{35Q20; 47J35.}

\makeatletter
\renewcommand{\@evenhead}{
\hspace*{-3pt}\raisebox{-7pt}[\headheight][0pt]{\vbox{\hbox to \textwidth {\thepage \hfil \evenhead}\vskip4pt \hrule}}}
\renewcommand{\@oddhead}{
\hspace*{-3pt}\raisebox{-7pt}[\headheight][0pt]{\vbox{\hbox to \textwidth {\oddhead \hfil \thepage}\vskip4pt\hrule}}}
\renewcommand{\@evenfoot}{}
\renewcommand{\@oddfoot}{}
\makeatother

\newpage
\vphantom{math}

\textcolor{blue!50!black}{\tableofcontents}
\vspace{0.7cm}

\textcolor{blue!50!black}{\section{A brief chronology of 150 years of the theory of kinetic equations}}

Mathematically the collective properties of many-particle systems in kinetic theory are described
by kinetic equations, namely, nonlinear equations that describe the evolution of the state of a system
of many particles by means of the evolution of the state of a typical particle. The generator of such
evolution equations consists of a term that describes the free-evolution (motion by inertia) of a typical
particle of the system and a nonlinear term that simulates the self-interaction of a typical particle
(collision integral). The mean values of observables of the many-particle system are determined by a
solution of the kinetic equations. A well-known historical example of a kinetic equation is the Boltzmann
equation, which describes the process of collision of particles in rarefied gases. Quantum kinetic equations
are the corresponding generalizations of the kinetic equations of classical particle systems.

From a physical point of view, kinetic equations describe a certain stage of the process of transition
(relaxation) from a non-equilibrium state to the state of thermodynamic equilibrium of the system of many
particles. Indeed, in the general case, in the process of relaxation, an arbitrary non-equilibrium state
of the particle system attracts to a state that can be fully described in terms of a one-particle probability
distribution function (for quantum systems, it is a one-particle statistical operator, the kernel of which is
known as a one-particle density matrix), that is governed by the suitable kinetic equation depending on the
interaction potential of the particles. At the next stage of relaxation, such a state of the particle system
tends to a state of local equilibrium, which is described by the Maxwell distribution, characterized by
hydrodynamic parameters that depend on a position in space. The evolution of hydrodynamic parameters in turn
is governed by the equations of a continuous medium (hydrodynamics or diffusion equations).

One hundred and fifty-year history of kinetic equations is derived from the works of J.~C.~Maxwell
(1860, 1867) \cite{Max60},\cite{Max67} and L.~E.~Boltzmann (1872) \cite{B72}. In 1872 Ludwig~Boltzmann published
the paper \cite{B72} where the evolution equation for a one-particle distribution function was formulated and was
showed that the Maxwell distribution describes only the equilibrium state of a gas. He proved the so-called H-theorem
(on the increase in entropy) about a property of the solution of this equation, which explained the irreversibility
of macroscopic dynamics. Thus, from this time began a period of development of the theory of kinetic equations,
which was based on phenomenological models of kinetic phenomena.

Later in order to generalize the Boltzmann equation to dense gases or liquids for a system of many hard spheres,
the Enskog kinetic equation was formulated (D.~Enskog, 1922) \cite{E22}. For a  particle (Brownian) in a system
of many particles (an environment) was introduced the Fokker--Planck equation (A.~D.~Fokker, 1914, M.~Planck, 1917)
\cite{F14},\cite{P17} and the Smoluchowski equation as its partial case (M.~von~Smoluchowski, 1906) \cite{S06}.
In this period of development of kinetic theory were also formulated: the Leontovich equation for stochastic dynamics
of a system of many particles (M.~A.~Leontovich, 1935) \cite{L35}, the Landau equation (L.~D.~Landau, 1936) \cite{L36},
the Vlasov equation (A.~A.~Vlasov, 1938) \cite{V38} and the Lenard--Balescu equation (A.~Lenard, R.~Balescu, R.~L.~Guernsey, 1960)
\cite{B60},\cite{G62},\cite{L60} for systems of many charged particles (ionized gases, plasma).

With the beginning of the development of quantum theory the Uehling--Uhlenbeck kinetic equation was formulated
as a quantum analog of the Boltzmann equation (L.~W.~Nordheim, 1928; E.~A.~Uehling, G.~E.~Uhlenbeck, 1933) \cite{N28},\cite{UU33}
and the quantum Bogolyubov equation (M.~M.~Bogolyubov,  K.~P.~Gurov, 1947) \cite{BG47}.
In the mean-field approximation for pure states, the Hartree equation (D.~R.~Hartree, 1928) \cite{H28} or the Hartree--Fock
equation for systems of fermions and bosons (V.~A.~Fock, 1930) \cite{F30} were derived, and for mixed states, the quantum
Vlasov kinetic equation (A.~A.~Vlasov, 1947) \cite{V61}. For quantum many-particle systems in condensed states, it was
suggested the Bogolyubov kinetic equation (M.~M.~Bogolyubov, 1947) \cite{B47} and later the Gross--Pitaevskii equation
(E.~P.~Gross, L.~D.~Pitaevskii, 1961) \cite{G61},\cite{P61}.

From the second half of the 1940s, a new stage in the development of kinetic theory began, namely, the creation
of a formalized theory of kinetic phenomena. In 1945 in Kyiv at the Institute of Mathematics in the famous
monograph \cite{Bog}, M.~M.~Bogolyubov formulated a consistent approach to deriving kinetic equations from the
dynamics of systems of many particles, namely, from fundamental equations described the evolution of all possible
states of many-particle systems, they are now known as the BBGKY hierarchy (Bogolyubov--Born--Green--Kirkwood--Yvon)
\cite{Bog},\cite{BG46},\cite{K47},\cite{Y35}. Using the methods of perturbation theory, an approach was developed
to construct a generalization of the Boltzmann equation, known as the kinetic equation of Bogolyubov, as well as
the Vlasov and Landau kinetic equations. Thanks to this work, the irreversibility mechanism of the evolution
of systems of many particles in the macroscopic scale, the dynamics of which are described reversible in time
equations of motion in a microscopic scale, became clear. Initially, the mathematical theory of the BBGKY hierarchy
was developed in works \cite{GP85},\cite{Pe71},\cite{GP83} and see also in \cite{CGP97},\cite{G12},\cite{Pe95}.

Later H.~Grad (H.~Grad, 1958) \cite{G58} formulated an approach to the derivation of kinetic equations as the
evolution equations, which describe the corresponding scaling asymptotic of a solution of the BBGKY hierarchy.
In recent decades, this approach has been used as an accepted method for the rigorous derivation of kinetic
equations of complex systems of various nature. In general, the problem of the rigorous derivation of kinetic
equations from the dynamics of systems of many particles remains an open problem of kinetic theory.

At the present stage of development of kinetic theory, the most advanced is the mathematical theory of the
nonlinear Boltzmann equation, which takes its origin from the works of H.~Poincar\'{e} (1906) \cite{P06},
who drew the attention of mathematicians to the need to substantiate kinetic theory and D.~Hilbert (1912) \cite{H12},
who established the connection between a solution of the Boltzmann equation and the hydrodynamics equations of
(D.~Hilbert's sixth problem was formulated at the International Congress of Mathematics in 1900 \cite{H02}),
as well as the work of T.~Carleman (1957) \cite{C57} on mathematical analysis of the spatially homogeneous
Boltzmann equation.

The mathematical theory of nonlinear kinetic equations began to develop intensively in the early 80s of the XX
century. One of the achievements of this period was the rigorous derivation of the Boltzmann equation from the
dynamics of an infinite number of hard spheres in the Boltzmann--Grad limit \cite{L75},\cite{PG}. Rigorous
results of the theory of kinetic equations and their justification at the end of the XX century were summarized
in monographs \cite{CGP97},\cite{CIP},\cite{Sp91}.

Over the last decade mathematical results on the derivation of the Boltzmann kinetic equation for classical
systems of particles with short-range interaction potential have been summarized in the monograph \cite{SR12}
and new methods have been developed for deriving the Boltzmann equation \cite{GG18} and the Enskog equation
\cite{GG12} from the collisional dynamics \cite{GG21} (for details see the references in the above works).

In the last two decades significant progress has also been observed in deriving quantum kinetic equations
in the scaling limits of the BBCKY hierarchy solution constructed by methods of the perturbation theory
\cite{BPS}, \cite{G13}, \cite{Pe95}, in particular, of the quantum Boltzmann equation \cite{BCEP3} in a weak
coupling limit, of the nonlinear Schr\"{o}dinger equation \cite{BSh17}, \cite{PP09} and the Gross--Pitaevskii
equation \cite{EShY10} in the mean-field limit.

Below the fundamental equations that describe the nature of substance are studies, namely the hierarchies of
evolution equations of quantum systems of many particles, and non-perturbative methods for constructing their
solutions. Based on this, two new approaches to the description of the kinetic evolution of quantum systems
\cite{G12} are considered. One of them is consists in the description of the evolution of quantum systems
in the mean-field scaling limit of the hierarchy of evolution equations for the observables \cite{G11} and
an approach based on a non-Markovian generalization of quantum kinetic equations \cite{GT}.

\textcolor{blue!50!black}{\section{Hierarchies of evolution equations of quantum systems}}

As is well-known, quantum systems are described by concepts such as observables and states. The average value
functional of the observables (mathematical expectation) determines the duality of the observables and the state.
Consequently, there are two approaches to describing the evolution of a quantum system of a finite number of
particles, namely, in terms of observables whose evolution is governed by the Heisenberg equation, or in terms
of a state whose evolution is governed by the von Neumann equation (quantum Liouville equation) for the density
operator (statistical operator) which kernel is known as the density matrix, respectively.

An equivalent approach to describing the evolution of systems of many particles, both of a finite and an infinite
number of particles, is to describe the evolution in terms of a sequence of reduced operators of observables are
governing by the dual BBGKY hierarchy (Bogolyubov--Born--Green--Kirkwood--Yvon) \cite{G12},\cite{GB}, or in terms
of the state describing by a sequence of the reduced density operators governed by the BBGKY hierarchy \cite{Bog}.

An alternative method for describing the evolution of the state of a quantum system of finitely many particles is
to describe the state in terms of the operators that are determined by cluster expansions of the density operator.
Such operators are interpreted as correlations of particle states, and their evolution is governed by the von Neumann
hierarchy for a sequence of correlation operators \cite{G19},\cite{GP},\cite{GerShJ}.

\textcolor{blue!50!black}{\subsection{Preliminaries: on evolution equations of quantum systems of many particles}}
For generality, we will consider below quantum systems in space $\mathbb{R}^{\nu},\,\nu\geq1$of a non-fixed
number of identical spinless particles, i.e. of an arbitrary but finite average number of particles, which
obeying the Maxwell--Boltzmann statistics. We will use units where $h={2\pi\hbar}=1$ is a Planck constant,
$m =1$ is particle mass.

We denote the $n$-particle Hilbert space by $\mathcal{H}_n=\mathcal{H}^{\otimes n}$ and
$\mathcal{H}^{\otimes 0}=\mathbb{C}$.  We denote the Fock space over the space $\mathcal{H}$ by
$\mathcal{F}_{\mathcal{H}}={\bigoplus\limits}_{n=0}^{\infty}\mathcal{H}_{n}$. The self-adjoint operator
$f_{n}$ defined on the space $\mathcal{H}_{n}=\mathcal{H}^{\otimes n}$ further we will also denote by
the symbol $f_{n}(1,\ldots,n)$. Let $\mathfrak{L}(\mathcal{H}_{n})$ be a space of bounded operators
$g_{n}\equiv g_{n}(1,\ldots,n)\in\mathfrak{L}(\mathcal{H}_{n})$ with operator norm
$\|.\|_{\mathfrak{L}(\mathcal{H}_{n})}$.
Accordingly, let the space $\mathfrak{L}^{1}(\mathcal{H}_{n})$ be the space of trace class operators
$f_{n}\equiv f_{n}(1,\ldots,n)\in\mathfrak{L}^{1}(\mathcal{H}_{n})$ with the norm:
$\|f_{n}\|_{\mathfrak{L}^{1}(\mathcal{H}_{n})}=\mathrm{Tr}_{1,\ldots,n}|f_{n}(1,\ldots,n)|,$
where the symbol $\mathrm{Tr}_{1,\ldots,n}$ denotes partial traces of the operator $f_{n}$. The subspace
of finite sequences of degenerate operators with infinitely differentiated kernels with compact supports
is denoted by $\mathfrak{L}^{1}_0(\mathcal{F}_{\mathcal{H}})$.

For a quantum system of a non-fixed number of particles, the observables are described by sequences
$A=(A_0,A_{1}(1),\ldots,A_{n}(1,\ldots,n),\ldots)$ of the self-adjoint operators
$A_{n}\in\mathfrak{L}(\mathcal{H}_{n})$. The evolution of the observables
$A(t)=(A_0,A_{1}(t,1),\ldots,A_{n}(t,1,\ldots,n),\ldots)$, where $t\in\mathbb{R}$,  is determined by
the Cauchy problem for a sequence of the Heisenberg equations:
\begin{eqnarray}
  \label{H-N1}
     &&\frac{\partial}{\partial t}A(t)=\mathcal{N}A(t),\\
  \label{H-N12}
     &&A(t)|_{t=0}=A(0),
\end{eqnarray}
where $A(0)=(A_0,A_{1}^0(1),\ldots,A_{n}^0(1,\ldots,n),\ldots)$ is the initial observable, the generator
$\mathcal{N}=\oplus_{n=0}^\infty\mathcal{N}_n$ is defined by the formula
\begin{eqnarray}\label{info}
    &&\mathcal{N}_n g_n\doteq-i\,(g_n H_n - H_n g_n ),
\end{eqnarray}
and the self-adjoint operator
$H_{n}={\sum\limits}_{j=1}^{n}K(j)+\epsilon{\sum\limits}_{j_{1}<j_{2}=1}^{n}\Phi(j_{1},j_{2})$ is the
Hamilton operator of a system of $n$ particles, that is, the operator $K(j)$ is the kinetic energy operator
of $j$ particle, $\Phi$ is the bounded operator of the pair interaction potential, $\epsilon>0$ is the
scaling parameter.

If $A(0)\in\mathfrak{L}(\mathcal{F}_{\mathcal{H}})$, then for $t\in\mathbb{R}$ in the space
$\mathfrak{L}(\mathcal{F}_{\mathcal{H}})$ the unique solution of the Cauchy problem
\eqref{H-N1},\eqref{H-N12} is represented by a one-parameter family of mappings
$\mathcal{G}(t)=\oplus_{n=0}^\infty\mathcal{G}_n(t)$:
\begin{eqnarray*}\label{sH}
    &&A(t)=\mathcal{G}(t)A(0),
\end{eqnarray*}
where the mapping $\mathcal{G}_n(t)$ is defined by the formula
\begin{eqnarray}\label{grG}
    &&\mathbb{R}^1\ni t\mapsto\mathcal{G}_n(t)g_n\doteq e^{itH_{n}}g_n e^{-itH_{n}}.
\end{eqnarray}
In the space $\mathfrak{L} (\mathcal{H}_{n})$ one-parameter mapping (\ref{grG}) forms a $\ast$-weakly
continuous group of operators, the infinitesimal generator of which coincides with operator
\eqref{info} on the domain of its definition.

The average value (mathematical expectation) of the observable $A(t)$ at time $t\in\mathbb{R}$
is determined by a continuous linear functional, which is represented by such an expansion in
a series \cite{G12}:
\begin{eqnarray}\label{averageD}
     &&\langle A\rangle(t)=(I,D(0))^{-1}\sum\limits_{n=0}^{\infty}\frac{1}{n!}
         \,\mathrm{Tr}_{1,\ldots,n}\,A_{n}(t)\,D_{n}^0,
\end{eqnarray}
where the sequence $D(0)=(I,D_{1}^0,\ldots,D_{n}^0,\ldots)$ of positive self-adjoint operators
$D_{n}^0\in\mathfrak{L}^1(\mathcal{H}_{n})$ is the sequence of density operators by which they
are described all possible states of the quantum system of a non-fixed number of particles at
the initial instant, the coefficient
$$(I,D(0))={\sum\limits}_{n=0}^{\infty}\frac{1}{n!}\mathrm{Tr}_{1,\ldots,n}D_{n}^0$$
is the normalizing factor. Functional \eqref{averageD}, which determines the duality of observables
and a state, exists for $D_{n}^0\in\mathfrak{L}^{1}(\mathcal{H}_{n})$ and
$A_{n}(t)\in\mathfrak{L}(\mathcal{H}_{n})$.

For the normalizing factor, the following equality is valid
\begin{eqnarray}\label{nor}
     &&(I,D(0))=(I,\mathcal{G}^\ast(t)D(0)).
\end{eqnarray}
On the space $\mathfrak{L}^1(\mathcal{F}_{\mathcal{H}})$ the one-parameter mapping
$\mathcal{G}^\ast(t)=\oplus_{n=0}^\infty \mathcal{G}^\ast_n(t)$ adjoint to the group \eqref{grG}
is defined by
\begin{eqnarray}\label{grGs}
    &&\mathbb{R}^1\ni t\mapsto\mathcal{G}^{\ast}_n(t)f_n\doteq e^{-itH_{n}}f_n e^{itH_{n}},
\end{eqnarray}
and forms a strongly continuous isometric group of operators, which preserves positivity and
self-adjointness of operators.

Due to equality \eqref{nor} for functional \eqref{averageD} the following representation is valid:
\begin{eqnarray*}
     &&\hskip-12mm (A(t),D(0))=(I,D(0))^{-1}\sum\limits_{n=0}^{\infty}\frac{1}{n!}
         \,\mathrm{Tr}_{1,\ldots,n}\,\mathcal{G}_{n}(t)A_{n}^0\,D_{n}^0=\\
     &&(I,\mathcal{G}^\ast(t)D(0))^{-1}\sum\limits_{n=0}^{\infty}\frac{1}{n!}\,
         \mathrm{Tr}_{1,\ldots,n}\,A_{n}^0\,\mathcal{G}^\ast_n(t)D_{n}^0=\\
     &&(I,D(t))^{-1}(A(0),D(t)),
\end{eqnarray*}
that is, the evolution of quantum systems of many particles in an equivalent way can be described as
the evolution of the state. Indeed, the evolution of all possible states, i.e. the sequence
$D(t)=(1,D_{1}(t),\ldots,D_{n}(t),\ldots)\in \mathfrak{L}^{1}(\mathcal{F}_{\mathcal{H}})$
of the density operators $D_{n}(t),\,n\geq1$, is described by the Cauchy problem for a sequence of the
von Neumann equations (quantum Liouville equations):
\begin{eqnarray}
  \label{vonNeumannEqn}
     &&\frac{\partial}{\partial t}D(t)=\mathcal{N}^\ast D(t),\\
  \label{F-N12}
     &&D(t)|_{t=0}=D(0),
\end{eqnarray}
where the generator $\mathcal{N}^\ast=\oplus^{\infty}_{n=0}\mathcal{N}^\ast_{n}$ of the von Neumann
equation \eqref{vonNeumannEqn} is adjoint operator to generator \eqref{info} of the Heisenberg equation
\eqref{H-N1} and is defined by the formula
\begin{eqnarray}\label{infOper1}
    &&\mathcal{N}^{\ast}_n f_n\doteq-i\,(H_n f_n - f_n H_n).
\end{eqnarray}
Operator \eqref{infOper1} has such structure:
$\mathcal{N}^{\ast}_n=\sum_{j=1}^{n}\mathcal{N}^{\ast}(j)+
\epsilon\sum_{j_{1}<j_{2}=1}^{n}\mathcal{N}^{\ast}_{\mathrm{int}}(j_{1},j_{2})$,
where the operator $\mathcal{N}^{\ast}(j)$ is the generator of the von Neumann equation of noninteracting
particles and the operator $\mathcal{N}^{\ast}_{\mathrm{int}}$ is defined is defined by the operator of the
particle interaction:
$\mathcal{N}^{\ast}_{\mathrm{int}}(j_{1},j_{2})f_n\doteq -i\,(\Phi(j_{1},j_{2})f_n-f_n\Phi(j_{1},j_{2}))$.

Thus, the unique solution of the Cauchy problem \eqref{vonNeumannEqn},\eqref{F-N12} is representing by the
group of operators \eqref{grG}
\begin{eqnarray*}\label{rozv_fon-N}
    &&D(t)=\mathcal{G}^\ast(t)D(0).
\end{eqnarray*}

Note that the density operator is represented by a convex linear combination of projectors of the first rank.
Density operator, which is a projector of the first rank $D_{n}(t)=P_{\psi_n}(t),\,\psi_n\in\mathcal{H}_{n},$
is known as the pure state, and the arbitrary state is interpreted as a mixed state. As a consequence of the
validity for the projector of such equality: $P_{\psi_n}(t)=P_{\psi_n(t)}$, where $\psi_n(t)=e^{-itH_{n}}\psi_n$,
the evolution of the pure state can also be described by the Cauchy problem for a sequence of the Schr\"{o}dinger
equations:
\begin{eqnarray*}
  \label{Sch}
     &&i\frac{\partial}{\partial t}\psi_{n}(t)=H_{n}\psi_{n}(t),\\
  \label{Schi}
     &&\psi(t)_{n}\big|_{t=0}=\psi_n^0, \quad n\geq1,
\end{eqnarray*}
where the operator $H_{n}$ is the Hamiltonian of the system of $n$ particles.

\textcolor{blue!50!black}{\subsection{Dynamics of correlations of particle states}}
An alternative approach to describing the state of a quantum system of a finite average number of
particles is to describe the state using cumulants of density operators, which are interpreted as
correlations of the states of clusters of particles \cite{G199},\cite{GP},\cite{GerShJ}.

We introduce sequence $g(t)=(I,g_{1}(t,1),\ldots,g_{s}(t,1,\ldots,s),\ldots)$ of correlation operators
by means of cluster expansions of the density operators $D(t)=(I,D_{1}(t,1),\ldots,D_{n}(t,1,\ldots,n),\ldots)$:
\begin{eqnarray}\label{D_(g)N}
    &&\hskip-12mm D_{n}(t,1,\ldots,n)= g_{n}(t,1,\ldots,n)+\sum\limits_{\mbox{\scriptsize $\begin{array}{c}\mathrm{P}:
    (1,\ldots,n)=\bigcup_{i}X_{i},\\|\mathrm{P}|>1 \end{array}$}}
        \prod_{X_i\subset \mathrm{P}}g_{|X_i|}(t,X_i),\quad n\geq1,
\end{eqnarray}
where ${\sum\limits}_{\mathrm{P}:(1,\ldots,n)=\bigcup_{i} X_{i},\,|\mathrm{P}|>1}$ is the sum over all
possible partitions $\mathrm{P}$ of the set of indices $(1,\ldots,n)$ on $|\mathrm{P}|>1$ non-empty subsets
$X_i\subset (1,\ldots,n)$, which are not mutually intersecting.

The solutions of recurrent relations \eqref{D_(g)N} are determined by the following expansions:
\begin{eqnarray}\label{gfromDFB}
   &&\hskip-15mm g_{s}(t,1,\ldots,s)=D_{s}(t,1,\ldots,s)+\\
   &&\hskip-12mm \sum\limits_{\mbox{\scriptsize $\begin{array}{c}\mathrm{P}:(1,\ldots,s)=
       \bigcup_{i}X_{i},\\|\mathrm{P}|>1\end{array}$}}(-1)^{|\mathrm{P}|-1}(|\mathrm{P}|-1)!\,
       \prod_{X_i\subset \mathrm{P}}D_{|X_i|}(t,X_i), \quad s\geq1. \nonumber
\end{eqnarray}
The structure of expansions \eqref{gfromDFB} is such that correlation operators can be interpreted
as cumulants (semi-invariants) of density operators \eqref{rozv_fon-N}.

If $g_{s}^{0}\in\mathfrak{L}^{1}(\mathcal{H}_{s}),\, s\geq1$, then for $t\in\mathbb{R}$ sequence of
correlation operators \eqref{gfromDFB} is the unique solution of the Cauchy problem of the von Neumann
hierarchy \cite{GP},\cite{GerShJ}:
\begin{eqnarray}\label{vNh}
   &&\hskip-15mm \frac{\partial}{\partial t}g_{s}(t,1,\ldots,s)=\mathcal{N}^{\ast}_{s}g_{s}(t,1,\ldots,s)+\\
   &&\hskip-12mm \sum\limits_{\mathrm{P}:\,(1,\ldots,s)=X_{1}\bigcup X_2}\,\sum\limits_{i_{1}\in X_{1}}
      \sum\limits_{i_{2}\in X_{2}}\epsilon\, \mathcal{N}_{\mathrm{int}}^{\ast}(i_{1},i_{2})
      g_{|X_{1}|}(t,X_{1})g_{|X_{2}|}(t,X_{2}), \nonumber\\
   \nonumber\\
 \label{vNhi}
   &&\hskip-15mm g_{s}(t,1,\ldots,s)\big|_{t=0}=g_{s}^{0}(1,\ldots,s),\quad s\geq1,
\end{eqnarray}
where the symbol ${\sum\limits}_{\mathrm{P}:\,(1,\ldots,s)=X_{1}\bigcup X_2}$ means the sum over all
possible partitions $\mathrm{P}$ of the set $(1,\ldots,s)$ on two nonempty subsets $X_1$ and $X_2$, which
are not mutually intersect, and the operators $\mathcal{N}^{\ast}_{s}$, $\mathcal{N}_{\mathrm{int}}^{\ast}(i_{1},i_{2})$
are defined by formulas \eqref{infOper1}. We emphasize that the von Neumann hierarchy \eqref{vNh} is a set
of recurrent evolution equations.

If the initial state is described by a sequence of correlation operators
$g(0)=(I,g_{1}^{0}(1),\ldots,$ $g_{n}^{0}(1,\ldots,n),\ldots)\in\oplus_{n=0}^{\infty}\mathfrak{L}^{1}(\mathcal{H}_{n})$,
then the evolution of all possible states of a quantum system of many particles, i.e., the sequence
$g(t)=(I,g_{1}(t,1),\ldots,g_{s}(t,1,\ldots,s),\ldots)$ of correlation operators $g_{s}(t),\,s\geq1$,
is determined by such a group of nonlinear operators \cite{GP}:
\begin{eqnarray}\label{ghs}
   &&\hskip-15mm g_{s}(t,1,\ldots,s)=\mathcal{G}(t;1,\ldots,s\mid g(0))\doteq\\
   &&\hskip-10mm \sum\limits_{\mathrm{P}:\,(1,\ldots,s)=\bigcup_j X_j}
      \mathfrak{A}_{|\mathrm{P}|}(t,\{X_1\},\ldots,\{X_{|\mathrm{P}|}\})
      \prod_{X_j\subset \mathrm{P}}g_{|X_j|}^{0}(X_j),\quad s\geq1,\nonumber
\end{eqnarray}
where $\sum_{\mathrm{P}:\,(1,\ldots,s)=\bigcup_j X_j}$ is the sum over all possible partitions
$\mathrm{P}$ of the set of indices $(1,\ldots,n)$ on $|\mathrm{P}|>1$ non-empty subsets $X_j$,
which are not mutually intersecting, the set $(\{X_1\},\ldots,\{X_{|\mathrm{P}|}\})$ consists
of elements that are subsets $X_j\subset (1,\ldots,s)$, i.e.,
$|(\{X_1\},\ldots,\{X_{|\mathrm{P}|}\})|=|\mathrm{P}|$. The generating operator
$\mathfrak{A}_{|\mathrm{P}|}(t)$ of expansion \eqref{ghs} is the $|\mathrm{P}|$-th order cumulant
of groups of operators \eqref{grGs}, which is determined by such expansion
\begin{eqnarray}\label{cumulantP}
   &&\hskip-15mm \mathfrak{A}_{|\mathrm{P}|}(t,\{X_1\},\ldots,\{X_{|\mathrm{P}|}\})\doteq\\
   &&\hskip-5mm \sum\limits_{\mathrm{P}^{'}:\,(\{X_1\},\ldots,\{X_{|\mathrm{P}|}\})=
      \bigcup_k Z_k}(-1)^{|\mathrm{P}^{'}|-1}({|\mathrm{P}^{'}|-1})!
      \prod\limits_{Z_k\subset\mathrm{P}^{'}}\mathcal{G}^{\ast}_{|\theta(Z_{k})|}(t,\theta(Z_{k})),\nonumber
\end{eqnarray}
where $\theta$ is the declusterization mapping: $\theta(\{X_1\},\ldots,\{X_{|\mathrm{P}|}\})\doteq(1,\ldots,s)$.

In the absence of correlations between the particles at the initial instant (the initial state satisfies the chaos
condition \cite{BPS},\cite{Sp91}), that is, of the sequence of initial correlation operators
$g^{(c)}(0)=(0,g_{1}^{0}(1),0,\ldots,0,\ldots)$ (in the case of the Maxwell--Boltzmann statistics in terms
of density operators this condition means that $D^{(c)}(0)=(I,D_{1}^0(1),\ldots,\prod^n_{i=1}D_{1}^0(i),\ldots)$),
expansions \eqref{ghs} take the form:
\begin{eqnarray*}\label{gth}
   &&g_{s}(t,1,\ldots,s)=\mathfrak{A}_{s}(t,1,\ldots,s)\,\prod\limits_{i=1}^{s}g_{1}^{0}(i),\quad s\geq1,
\end{eqnarray*}
where the $s$-th order cumulant $\mathfrak{A}_{s}(t)$ of the groups of operators \eqref{grGs} is defined
by the following expansion
\begin{eqnarray}\label{cumcp}
   &&\hskip-12mm \mathfrak{A}_{s}(t,1,\ldots,s)=\sum\limits_{\mathrm{P}:\,(1,\ldots,s)=
       \bigcup_i X_i}(-1)^{|\mathrm{P}|-1}({|\mathrm{P}|-1})!
      \prod\limits_{X_i\subset\mathrm{P}}\mathcal{G}^{\ast}_{|X_i|}(t,X_i),
\end{eqnarray}
and where it was used notations accepted in formula \eqref{ghs}.

Thus, the cumulant origin of correlation operators \eqref{gfromDFB} induces a cumulant structure
of one-parameter mapping (\ref{ghs}).

We emphasize that the dynamics of correlations, i.e. the hierarchy of fundamental equations (\ref{vNh})
which describes the evolution of state correlations can be used as a basis for describing the evolution
of states both a system of a finite and infinite number of particles instead of the von Neumann equations
\eqref{vonNeumannEqn} for density operators.

\textcolor{blue!50!black}{\subsection{The BBGKY hierarchies of evolution equations}}
To describe the evolution of quantum systems of both finite and infinite numbers of particles, another
approach to the description of observables and a state is used, which equivalent to the approach formulated
above in the case of systems of a finite average number of particles. \cite{BogLect},\cite{CGP97}.

Indeed, the mean value functional of observables \eqref{averageD} can be else represented in the following form
\begin{eqnarray}\label{avmar}
       &&\hskip-12mm \langle A\rangle(t)=(I,D(0))^{-1}\sum\limits_{n=0}^{\infty}\frac{1}{n!}\,
         \mathrm{Tr}_{1,\ldots,n}\,A_{n}(t)\,D_{n}^0=\\
       && \sum\limits_{s=0}^{\infty}\frac{1}{s!}\,
         \mathrm{Tr}_{1,\ldots,s}\,B_{s}(t,1,\ldots,s)\,F_{s}^{0}(1,\ldots,s),\nonumber
\end{eqnarray}
where to describe the observables and the state introduced sequences of the reduced ($s$-particle) operators
of observables $B(t)=(B_0,B_{1}(t,1),\ldots,B_{s}(t,1,\ldots,s),\ldots)$ and the reduced ($s$-particle) density
operators $F(0)=(I,F_{1}^{0}(1),\ldots,F_{s}^{0}(1,\ldots,s),\ldots)$, respectively \cite{BogLect},\cite{Sp91}.
Therefore, the reduced observables are defined in terms of the observables by such expansions \cite{GB}:
\begin{eqnarray}\label{mo}
    &&\hskip-12mm B_{s}(t,1,\ldots,s)=\sum_{n=0}^s\,\frac{(-1)^n}{n!}\sum_{j_1\neq\ldots\neq j_{n}=1}^s
          A_{s-n}(t,(1,\ldots,s)\setminus(j_1,\ldots,j_{n})), \quad s\geq 1,
\end{eqnarray}
and reduced density operators are defined by density operators as follows \cite{CGP97}:
\begin{eqnarray}\label{ms}
     &&\hskip-12mm F_{s}^{0}(1,\ldots,s)=(I,D)^{-1}\sum\limits_{n=0}^{\infty}\frac{1}{n!}\,
         \mathrm{Tr}_{s+1,\ldots,s+n}\,D_{s+n}^{0}(1,\ldots,s+n),\quad s\geq 1.
\end{eqnarray}

We emphasize that the possibility of describing observables and a state with the help of corresponding
reduced operators naturally arises as a result of dividing the series in the expression \eqref{averageD}
by the series of the normalization factor, i.e. in consequence of redefining the representation of mean
value functional \eqref{avmar}.

If at the initial moment an observable is determined by the sequence of reduced observables
$B(0)=(B_0,B_1^{0}(1),\ldots,B_s^{0}(1,\ldots,s),\ldots)\in\mathfrak{L}(\mathcal{F}_{\mathcal{H}})$,
then for arbitrary $t\in\mathbb{R}$ the sequence $B(t)=(B_0,B_1(t,1),\ldots,B_s(t,1,\ldots,s),\ldots)$
of reduced observables \eqref{mo} satisfies the Cauchy problem of the quantum dual BBGKY hierarchy
\cite{GShZ},\cite{G12},\cite{G11}:
\begin{eqnarray}\label{dh}
   &&\hskip-12mm \frac{\partial}{\partial t}B_{s}(t,1,\ldots,s)=\big(\sum\limits_{j=1}^{s}\mathcal{N}(j)+
      \epsilon\hskip-2mm\sum\limits_{j_1<j_{2}=1}^{s}\mathcal{N}_{\mathrm{int}}(j_1,j_{2})\big)B_{s}(t,1,\ldots,s)+\\
   &&\hskip-5mm+ \epsilon\,\sum_{j_1\neq j_{2}=1}^s
\mathcal{N}_{\mathrm{int}}(j_1,j_{2})B_{s-1}(t,1,\ldots,j_1-1,j_1+1,\ldots,s),\nonumber\\
      \nonumber\\
\label{dhi}
   &&\hskip-12mm B_{s}(t,1,\ldots,s)_{\mid t=0}=B_{s}^{0}(1,\ldots,s),\quad s\geq1,
\end{eqnarray}
where the notation accepted in formula \eqref{info} was used.

We remark that the hierarchy of equations (\ref{dh}) has the structure of recurrent evolution equations,
for example,
\begin{eqnarray*}
   &&\hskip-12mm\frac{\partial}{\partial t}B_{1}(t,1)=\mathcal{N}(1)B_{1}(t,1), \\
   &&\hskip-12mm\frac{\partial}{\partial t}B_{2}(t,1,2)=\big(\sum\limits_{j=1}^{2}\mathcal{N}(j)+
      \epsilon\,\mathcal{N}_{\mathrm{int}}(1,2)\big)B_{2}(t,1,2)+
      \epsilon\,\mathcal{N}_{\mathrm{int}}(1,2)\big(B_{1}(t,1)+B_{1}(t,2)\big).
\end{eqnarray*}

The solution of the Cauchy problem \eqref{dh},\eqref{dhi} is represented by the following expansions \cite{GB},\cite{G11}:
\begin{eqnarray}\label{sdh}
   &&\hskip-12mm B_{s}(t,1,\ldots,s)=\sum_{n=0}^s\,\frac{1}{n!}\sum_{j_1\neq\ldots\neq j_{n}=1}^s
      \mathfrak{A}_{1+n}\big(t,\{(1,\ldots,s)\setminus (j_1,\ldots,j_{n})\},\\
   &&(j_1,\ldots,j_{n})\big)\,
      B_{s-n}^{0}(1,\ldots,j_1-1,j_1+1,\ldots,j_n-1,j_n+1,\ldots,s),\quad s\geq1,\nonumber
\end{eqnarray}
where the generating operator of this expansion is the $(1+n)$-th order cumulant of the groups of operators \eqref{grG}:
\begin{eqnarray*}
    &&\hskip-12mm \mathfrak{A}_{1+n}(t,\{(1,\ldots,s)\setminus(j_1,\ldots,j_{n})\},(j_1,\ldots,j_{n}))\doteq\\
    &&\sum\limits_{\mathrm{P}:\,(\{(1,\ldots,s)\setminus(j_1,\ldots,j_{n})\},(j_1,\ldots,j_{n}))={\bigcup}_i X_i}
       (-1)^{\mathrm{|P|}-1}({\mathrm{|P|}-1})!\prod_{X_i\subset\mathrm{P}}
       \mathcal{G}_{|\theta(X_i)|}(t,\theta(X_i)),\quad  n\geq0,
\end{eqnarray*}
and used notation similar to the formula \eqref{cumulantP}.

We note that expansion (\ref{sdh}) for the solution can be represented as an iteration series
(perturbation theory series) of recurrent evolution equations \eqref{dh} in the result of the
application of analogues of the Duhamel equation to the generating operators, i.e. to the
cumulants of groups of operators \eqref{grG}.

According to the definition of additive type observables \eqref{mo} correspond to the one-component
sequences $B^{(1)}(0)=(0,b_{1}^{0}(1),0,\ldots)$ of reduced observables, and the sequences
$B^{(k)}(0)=(0,\ldots,$ $0,b_{k}^{0}(1,\ldots,k),0,\ldots)$ correspond to $k$-ary (non-additive) type
observables.
Then for a certain type of observables, the structure of expansion (\ref{sdh}) takes the appropriate form
\begin{eqnarray}\label{af}
     &&B_{s}^{(1)}(t,1,\ldots,s)=\mathfrak{A}_{s}(t,1,\ldots,s)\sum_{j=1}^s b_{1}^{0}(j), \quad s\geq 1,
\end{eqnarray}
where the generating operator $\mathfrak{A}_{s}(t)$ is the $s$-th order cumulant of the groups of operators
\eqref{grG}, and, if $s\geq k$,
\begin{eqnarray}\label{af-k}
     &&\hskip-12mm B_{s}^{(k)}(t,1,\ldots,s)=\frac{1}{(s-k)!}\sum_{j_1\neq\ldots\neq j_{s-k}=1}^s
      \mathfrak{A}_{1+s-k}\big(t,\{(1,\ldots,s)\setminus (j_1,\ldots,j_{s-k})\}, j_1,\ldots,\\
     && j_{s-k}\big)\,b_{k}^{0}(1,\ldots,j_1-1,j_1+1,\ldots,j_s-k-1,j_s-k+1,\ldots,s),\nonumber
\end{eqnarray}
and, if $1\leq s<k$, we have: $B_{s}^{(k)}(t)=0$.

Traditionally, the evolution of many-particle systems is described within the framework of the evolution
of the state governed by the BBGKY hierarchy for reduced density operators \cite{BPS},\cite{BogLect},\cite{CGP97},\cite{Sp91}.

Indeed, for functional \eqref{avmar} the following representation is hold
\begin{eqnarray*}
     &&(B(t),F(0))=(B(0),F(t)),
\end{eqnarray*}
that is, the evolution of quantum systems of many particles in an equivalent way can be described as
the evolution of the state using reduced density operators \eqref{ms}.

If $F(0)\in\mathfrak{L}^{1}(\mathcal{F}_{\mathcal{H}})$, the for arbitrary $t\in\mathbb{R}$
the sequence $F(t)=(I,F_1(t,1),\ldots,F_s(t,$ $1,\ldots,s),\ldots)$ of reduced density operators
(\ref{ms}) satisfies the Cauchy problem of the quantum BBGKY hierarchy \cite{BogLect},\cite{CGP97}:
\begin{eqnarray}
 \label{BBGKY}
   &&\hskip-12mm \frac{\partial}{\partial t}F_{s}(t,1,\ldots,s)=\mathcal{N}^{\ast}_{s}F_{s}(t,1,\ldots,s)+
       \epsilon\,\sum\limits_{j=1}^{s}\mathrm{Tr}_{s+1}\mathcal{N}^{\ast}_{\mathrm{int}}(j,s+1)F_{s+1}(t,1,\ldots,s,s+1),\\
       \nonumber \\
 \label{BBGKYi}
   &&\hskip-12mm F_{s}(t,1,\ldots,s)\mid_{t=0}=F_{s}^{0}(1,\ldots,s),\quad s\geq 1,
\end{eqnarray}
where the notation from formula \eqref{infOper1} is used.

The solution of the Cauchy problem \eqref{BBGKY},\eqref{BBGKYi} is represented by the following
series \cite{GerRS},\cite{GerS}:
\begin{eqnarray}\label{RozvBBGKY}
   &&\hskip-12mm F_{s}(t,1,\ldots,s)=\\
   &&\sum\limits_{n=0}^{\infty}\frac{1}{n!}\,\mathrm{Tr}_{s+1,\ldots,{s+n}}\,
       \mathfrak{A}_{1+n}(t,\{1,\ldots,s\},s+1,\ldots,{s+n})F_{s+n}^{0}(1,\ldots,{s+n}),\quad s\geq1,\nonumber
\end{eqnarray}
where the generating operator of this series
\begin{eqnarray}\label{cumulant1+n}
   &&\hskip-15mm \mathfrak{A}_{1+n}(t,\{1,\ldots,s\},s+1,\ldots,{s+n})=\\
   && \sum\limits_{\mathrm{P}\,:(\{1,\ldots,s\},s+1,\ldots,{s+n})=
      {\bigcup\limits}_i X_i}(-1)^{|\mathrm{P}|-1}(|\mathrm{P}|-1)!
      \prod_{X_i\subset\mathrm{P}}\mathcal{G}^{\ast}_{|\theta(X_i)|}(t,\theta(X_i))\nonumber
\end{eqnarray}
is the $(1+n)$-th order cumulant of the groups of operators \eqref{grGs}. In expansion \eqref{cumulant1+n}
symbol ${\sum\limits}_\mathrm{P}$ means the sum over all possible partitions $\mathrm{P}$ of the set
$(\{1,\ldots,s\},s+1,\ldots,{s+n})$ on $|\mathrm{P}|$ nonampty subsets
$X_i\subset(\{1,\ldots,s\},s+1,\ldots,{s+n})$, which do not mutually intersect,
and the notations introduced in \eqref{ghs} are used.

We remark that one of the methods for constructing solutions \eqref{RozvBBGKY} and \eqref{sdh} is based on
the application of cluster expansions \cite{G12},\cite{GerS} to groups of operators \eqref{grGs} and \eqref{grG},
which are the generating operators of the series \eqref{ms} for the reduced density operators and the expansions
\eqref{mo} for the reduced observables ones, respectively.

We note that a common form of representation of the solution of the Cauchy problem of the BBGKY hierarchy is its
representation as a perturbation theory series (iteration series of the BBGKY hierarchy) \cite{BogLect},\cite{Pe71}
(see also \cite{BPS} \cite{CGP97} and therein references):
\begin{eqnarray}\label{iter}
   &&\hskip-12mm F_s(t,1,\ldots,s)=\\
   &&\sum\limits_{n=0}^{\infty}\,\int\limits_{0}^{t}dt_{1}\ldots\int\limits_{0}^{t_{n-1}}dt_{n}
       \mathrm{Tr}_{s+1,\ldots,s+n}\mathcal{G}^{\ast}_s(t-t_{1})
       \sum\limits_{j_1=1}^{s}\mathcal{N}^{\ast}_{\mathrm{int}}(j_1,s+1))\mathcal{G}^{\ast}_{s+1}(t_1-t_2)\ldots\nonumber\\
   &&\mathcal{G}^{\ast}_{s+n-1}(t_{n-1}-t_n)\sum\limits_{j_n=1}^{s+n-1}\mathcal{N}^{\ast}_{\mathrm{int}}(j_n,s+n))
       \mathcal{G}^{\ast}_{s+n}(t_{n})F_{s+n}^0(1,\ldots,s+n), \quad s\geq1,\nonumber
\end{eqnarray}
where notations from the expression \eqref{infOper1} are used. The representation of the solution
by the series \eqref{RozvBBGKY} is equivalent to series \eqref{iter} due to the validity under
the appropriate conditions for the initial data and the interaction potential of particles of
analogs of the Duhamel equation for generating operators \eqref{cumulant1+n} of series
\eqref{RozvBBGKY} that is, for cumulants of groups of operators \eqref{grG}.

Thus, there are two approaches to describing the evolution of quantum systems of many particles, namely,
in terms of observables whose evolution is governed by the dual BBGKY hierarchy (\ref{dh}), or in terms
of a state whose evolution is governed by the BBGKY hierarchy (\ref{BBGKY}). For systems of finitely many
particles, these hierarchies of evolution equations are equivalent to the Heisenberg equation \eqref{H-N1}
and the von Neumann equation \eqref{vonNeumannEqn}, respectively.

In paper \cite{G12}, these hierarchies of evolution equations are generalized for many-particle systems with
multiparticle interaction potentials.

An alternative approach, as noted above, to the description of the evolution of the state of quantum systems,
consisting of both finitely and infinitely many particles, can be formulated by means of operators defined by
cluster expansions of the reduced density operators, namely:
\begin{eqnarray}\label{gBigfromDFB}
   &&\hskip-15mm G_{s}(t,1,\ldots,s)=
     \sum\limits_{\mbox{\scriptsize$\begin{array}{c}\mathrm{P}:(1,\ldots,s)=\bigcup_{i}X_{i}\end{array}$}}
     (-1)^{|\mathrm{P}|-1}(|\mathrm{P}|-1)!\,\prod_{X_i\subset \mathrm{P}}F_{|X_i|}(t,X_i), \quad s\geq1,
\end{eqnarray}
where the notation from formula \eqref{gfromDFB} was used. Such cumulants of reduced density operators are
interpreted as reduced  correlations of the state \cite{BogLect}.

On a microscopic scale, the macroscopic characteristics of the fluctuations of the observables are directly
determined by the reduced correlation operators, for example, the functional of dispersion of the additive
type observables, that is, of such sequences $A^{(1)}=(0,a_{1}(1),\ldots,\sum_{i=1}^{n}a_1(i),\ldots)$, is
represented by the following formula
\begin{eqnarray*}
    &&\hskip-12mm \langle(A^{(1)}-\langle A^{(1)}\rangle)^2\rangle(t)=
      \mathrm{Tr}_{1}\,(a_1^2(1)-\langle A^{(1)}\rangle^2(t))G_{1}(t,1)+
      \mathrm{Tr}_{1,2}\,a_{1}(1)a_{1}(2)G_{2}(t,1,2),
\end{eqnarray*}
where the mean value functional of observables of an additive type we denoted by
$\langle A^{(1)}\rangle(t)=\mathrm{Tr}_{1}\,a_{1}(1)G_{1}(t,1)$.

If $G(0)\in\in\mathfrak{L}^{1}(\mathcal{F}_{\mathcal{H}})$, then for arbitrary $t\in\mathbb{R}$
the sequence of reduced correlation operators \eqref{gBigfromDFB} satisfies the Cauchy problem for the hierarchy
of nonlinear evolution equations (quantun nonlinear BBGKY hierarchy) \cite{BogLect}:
\begin{eqnarray}\label{gBigfromDFBa}
   &&\hskip-12mm \frac{\partial}{\partial t}G_s(t,1,\ldots,s)=\mathcal{N}^{\ast}_{s}G_{s}(t,1,\ldots,s)+\\
   && \sum\limits_{\mathrm{P}:\,(1,\ldots,s)=X_{1}\bigcup X_2}\,\sum\limits_{i_{1}\in X_{1}}
      \sum\limits_{i_{2}\in X_{2}}\epsilon\,\mathcal{N}_{\mathrm{int}}^{\ast}(i_{1},i_{2})
      G_{|X_{1}|}(t,X_{1})G_{|X_{2}|}(t,X_{2}))+\nonumber\\
   && \mathrm{Tr}_{s+1}
      \sum_{i\in (1,\ldots,s)}\epsilon\,\mathcal{N}^{\ast}_{\mathrm{int}}(i,s+1)\big(G_{s+1}(t,1,\ldots,s+1)+\nonumber\\
    && \sum_{\mbox{\scriptsize$\begin{array}{c}\mathrm{P}:(1,\ldots,s+1)=X_1\bigcup X_2,\\i\in
      X_1;s+1\in X_2\end{array}$}}G_{|X_1|}(t,X_1)G_{|X_2|}(t,X_2)\big),\nonumber\\ \nonumber\\
 \label{gBigfromDFBai}
   &&\hskip-12mm G_{s}(t,1,\ldots,s)\big|_{t=0}=G_{s}^{0}(,1,\ldots,s), \quad s\geq1,
\end{eqnarray}
where the notations for the hierarchy of equations \eqref{vNh} is used.

In the case of the initial state in the absence of correlations between the particles, i.e. the state described by
the sequence $G^{(c)}=(0,G_1^{0},0,\ldots,0,\ldots)$, the solution of the Cauchy problem \eqref{gBigfromDFBa},\eqref{gBigfromDFBai}
is represented by the following series:
\begin{eqnarray}\label{mcc}
   &&\hskip-15mm G_{s}(t,1,\ldots,s)=\sum\limits_{n=0}^{\infty}\frac{1}{n!}
       \,\mathrm{Tr}_{s+1,\ldots,s+n}\,\mathfrak{A}_{s+n}(t;1,\ldots,s+n)
       \prod_{i=1}^{s+n}G_1^{0}(i), \quad s\geq1,
\end{eqnarray}
where the generating operator $\mathfrak{A}_{s+n}(t)$ is the $(s+n)$-th order cumulant \eqref{cumcp} of groups of
operators \eqref{grGs}.
We note that for the specified initial state, the expressions for the reduced density operators \eqref{RozvBBGKY}
and the reduced correlation operators \eqref{mcc} differ only in the order of the generating operator for the
corresponding terms of the series representing these operators.

For an arbitrary initial state \eqref{gBigfromDFBai} a solution of the Cauchy problem for the nonlinear BBGKY
hierarchy \eqref{gBigfromDFBa} was constructed in the paper \cite{G17} (see also \cite{G199}).

Note that the description of the evolution of the state of quantum systems of many particles by means of both
reduced density operators and reduced correlation operators can be based on an approach founded on the dynamics
of correlations, which is governed by the von Neumann hierarchy \eqref{vNh} for correlation operators. Within
the framework of this approach, the reduced density operators are defined by the following series:
\begin{eqnarray}\label{FClusters}
    &&\hskip-12mm F_{s}(t,1,\ldots,s)\doteq\sum\limits_{n=0}^{\infty}\frac{1}{n!}\,
       \mathrm{Tr}_{s+1,\ldots,s+n}\,\,g_{1+n}(t,\{1,\ldots,s\},s+1,\ldots,s+n), \quad s\geq1,
\end{eqnarray}
where the correlation operators $g_{1+n}(t),\,n\geq0,$ of particle cluster and particles are represented by the
following expansions:
\begin{eqnarray}\label{rozvNF-Nclusters}
    &&\hskip-12mm g_{1+n}(t,\{1,\ldots,s\},s+1,\ldots,s+n)=\\
    && \sum\limits_{\mathrm{P}:\,(\{1,\ldots,s\},\,s+1,\ldots,s+n)=\bigcup_i X_i}
       \mathfrak{A}_{|\mathrm{P}|}\big(t,\{\theta(X_1)\},\ldots,\{\theta(X_{|\mathrm{P}|})\}\big)
       \prod_{X_i\subset \mathrm{P}}g_{|X_i|}^0(X_i), \quad n\geq0,\nonumber
\end{eqnarray}
in which the notations from formula \eqref{ghs} are used. The reduced correlation operators are defined by the
corresponding series:
\begin{eqnarray}\label{GClusters}
   &&\hskip-8mm G_{s}(t,1,\ldots,s)\doteq\sum\limits_{n=0}^{\infty}\frac{1}{n!}\,
      \mathrm{Tr}_{s+1,\ldots,s+n}\,\,g_{s+n}(t,1,\ldots,s+n), \quad s\geq1,
\end{eqnarray}
where the correlation operators $g_{s+n}(t),\,n\geq0,$ are represented by expansions \eqref{ghs}.

Thus, as a result of definitions \eqref{FClusters} and \eqref{GClusters}, we establish that the cumulant
structure of generating operators of expansions for correlation operators \eqref{ghs}, or more general
expansions \eqref{rozvNF-Nclusters}, induces a cumulant structure of series for reduced density operators
\eqref{RozvBBGKY} and reduced correlation operators \eqref{mcc}, i.e. in fact, the evolution of a system
of infinitely many particles is generated by the dynamics of correlations.

Note also that in the article \cite{GF} the hierarchies of evolution equations
\eqref{vNh},\eqref{dh},\eqref{BBGKY},\eqref{gBigfromDFBa} are studied as evolution equations in functional derivatives.

\textcolor{blue!50!black}{\section{On quantum kinetic equations}}

This section describes an approach to describing the evolution of a state using the state of a typical
particle of a quantum system of many particles, or, in other words, discusses the origin of describing
the evolution of a state by means of quantum kinetic equations \cite{G09},\cite{PG98},\cite{GT}.

\textcolor{blue!50!black}{\subsection{The origin of the kinetic evolution of a state}}
We consider a system of many particles that satisfy the Maxwell-Boltzmann statistics, in the absence
of correlations between particles at the initial time, i.e. the initial state of which is determined
by a one-particle density operator, namely, such a sequence of reduced density operators
$F^{(c)}=(0,F_1^{0}(1),\ldots,\prod^n_{i=1}F_1^{0}(i),\ldots)$. We emphasize that the formulated
assumption about the initial state is inherent in the kinetic theory of systems of many particles
\cite{BPS},\cite{G12},\cite{G58}.

Due to the fact that the initial state is determined by a one-particle density operator, for the mean
value functional of observables \eqref{avmar} the following representation is valid
\begin{eqnarray*}\label{w}
    &&\big(B(t),F^{(c)}\big)=\big(B(0),F(t\mid F_{1}(t))\big),
\end{eqnarray*}
that is, the evolution of all possible states is described by a sequence
$F(t\mid F_{1}(t))=\big(I,F_1(t),F_2(t\mid F_{1}(t)),\ldots,F_s(t\mid F_{1}(t)),\ldots\big)$ of reduced
functionals of the state $F_{s}(t,1,\ldots,s\mid F_{1}(t)),\,s\geq2,$ which are represented by the series:
\begin{eqnarray}\label{f}
     &&\hskip-12mm F_{s}\bigl(t,1,\ldots,s\mid F_{1}(t)\bigr)=\\
     && \sum_{n=0}^{\infty }\frac{1}{n!}\,\mathrm{Tr}_{s+1,\ldots,s+n}\,
        \mathfrak{V}_{1+n}\bigl(t,\{1,\ldots,s\},s+1,\ldots,s+n\bigr)\prod_{i=1}^{s+n}F_{1}(t,i),\quad s\geq2,\nonumber
\end{eqnarray}
with respect of a one-particle density operator
\begin{eqnarray}\label{ske}
   &&\hskip-12mm F_{1}(t,1)=\sum\limits_{n=0}^{\infty}\frac{1}{n!}\,\mathrm{Tr}_{2,\ldots,{1+n}}\,
      \mathfrak{A}_{1+n}(t,1,\ldots,n+1)\prod_{i=1}^{n+1}F_{1}^{0}(i),
\end{eqnarray}
which describes the evolution of the state of a typical particle of a quantum system of many particles.
The generating operators of the series \eqref{ske} are the cumulants \eqref{cumulant1+n} of the
corresponding order of the groups of operators \eqref{grGs}.

The reduced functionals of the state  $F_s(t\mid F_{1}(t)),\,s\geq2$, describe all possible correlations
that created in the process of evolution of the quantum system of many particles in terms of the state
of a typical particle. The $(1+n)$-th order generating operator of the series \eqref{f} is determined by
the following expansion \cite{GT}
\begin{eqnarray}\label{skrrc}
   &&\hskip-9mm \mathfrak{V}_{1+n}\bigl(t,\{1,\ldots,s\},s+1,\ldots,s+n\bigr)=
                 n!\,\sum_{k=0}^{n}\,(-1)^k\,\sum_{n_1=1}^{n}\ldots\\
   &&\hskip-7mm \sum_{n_k=1}^{n-n_1-\ldots-n_{k-1}}\frac{1}{(n-n_1-\ldots-n_k)!}\,
               \widehat{\mathfrak{A}}_{1+n-n_1-\ldots-n_k}(t,\{1,\ldots,s\},s+1,\ldots, s+n-\nonumber\\
   &&\hskip-7mm n_1-\ldots-n_k)\prod_{j=1}^k\,\sum\limits_{\mbox{\scriptsize$\begin{array}{c}
       \mathrm{D}_{j}:Z_j=\bigcup_{l_j}X_{l_j},\\
       |\mathrm{D}_{j}|\leq s+n-n_1-\dots-n_j\end{array}$}}\frac{1}{|\mathrm{D}_{j}|!}
       \sum_{i_1\neq\ldots\neq i_{|\mathrm{D}_{j}|}=1}^{s+n-n_1-\ldots-n_j}\,
       \prod_{X_{l_j}\subset \mathrm{D}_{j}}\,\frac{1}{|X_{l_j}|!}\,
       \widehat{\mathfrak{A}}_{1+|X_{l_j}|}(t,i_{l_j},X_{l_j}),\nonumber
\end{eqnarray}
where the symbol $\sum_{\mathrm{D}_{j}:Z_j=\bigcup_{l_j} X_{l_j}}$ means the sum over all possible
dissections of the linearly ordered set $Z_j\equiv(s+n-n_1-\ldots-n_j+1,\ldots,s+n-n_1-\ldots-n_{j-1})$
on no more than $s+n-n_1-\ldots-n_j$ linearly ordered subsets and the generating operator of this expansion
$\widehat{\mathfrak{A}}_{1+n}(t)$ is the $(1+n)$-th order cumulant \eqref{cumulant1+n} of the groups of
scattering operators
\begin{eqnarray*}
    &&\widehat{\mathcal{G}}_{n}(t)\doteq
       \mathcal{G}_{n}^{\ast}(t,1,\ldots,n)\prod_{i=1}^{n}(\mathcal{G}_{1}^{\ast}(t,i))^{-1},\quad n\geq1,
\end{eqnarray*}
We give the simplest examples of generating operators of expansion \eqref{skrrc}:
\begin{eqnarray*}
   &&\hskip-8mm \mathfrak{V}_{1}(t,\{1,\ldots,s\})=\widehat{\mathfrak{A}}_{1}(t,\{1,\ldots,s\}),\\
   &&\hskip-8mm \mathfrak{V}_{1+1}(t,\{1,\ldots,s\},s+1)=\widehat{\mathfrak{A}}_{2}(t,\{1,\ldots,s\},s+1)-
               \widehat{\mathfrak{A}}_{1}(t,\{1,\ldots,s\})\sum_{i=1}^s\widehat{\mathfrak{A}}_{2}(t,i,s+1).
\end{eqnarray*}

Thus, according to the definition of \eqref{FClusters}, the cumulant structure of generating operators of
expansions for correlation operators \eqref{rozvNF-Nclusters} induces a generalized cumulant structure of
generating  operators for series of reduced functionals of the state \eqref{f}.

Noticing that for the initial state specified by the one-particle density operator $F^{(c)}$, the Cauchy
problem (\ref{gBigfromDFBa}),(\ref{gBigfromDFBai}) is not correctly well-defined, because the initial
data are not are independent for each unknown reduced density operator of the BBGKY hierarchy. As a
consequence, such a Cauchy problem can be reformulated as a new Cauchy problem for a one-particle density
operator with an independent initial condition and a sequence of explicitly defined functionals of the
solution of the Cauchy problem of the evolution equation for a one-particle density operator (kinetic
equation). In this case, the method of constructing reduced state functionals \eqref{f} is based on the
application of the species of cluster expansions, the so-called kinetic cluster expansions \cite{GT}, to
generating operators \eqref{cumulant1+n} of series representing reduced density operators \eqref{RozvBBGKY}.

\textcolor{blue!50!black}{\subsection{The generalized quantum kinetic equation}}
If $ F_{1}^{0}\in\mathfrak{L}^{1}(\mathcal{H})$, then for an arbitrary $t\in\mathbb{R}$ a one-particle
density operator \eqref{ske} satisfies the Cauchy problem for the generalized quantum kinetic equation \cite{GT}:
\begin{eqnarray}\label{gkec}
   &&\hskip-8mm\frac{\partial}{\partial t}F_{1}(t,1)=\mathcal{N}^{\ast}(1)F_{1}(t,1)+
        \epsilon\,\mathrm{Tr}_{2}\,\mathcal{N}_{\mathrm{int}}^{\ast}(1,2)F_{2}\bigl(t,1,2\mid F_{1}(t)\bigr),\\
   \nonumber\\
 \label{gkeci}
   &&\hskip-8mm F_{1}(t,1)\big|_{t=0}=F_{1}^{0}(1),
\end{eqnarray}
where the collision integral is determined by the two-particle functional of the state \eqref{f} and
the notation is used \eqref{infOper1}.

In \cite{GT}, the existence theorem of the Cauchy problem (\ref{gkec}), (\ref{gkeci}) in the space of
the trace class operators is proved.

Thus, for the initial state in the absence of correlations between particles, i.e. of the state specified
by the one-particle density operator, the evolution of all possible states of the quantum system of many
particles can be described without any approximations by means of a one-particle density operator (\ref{gkec})
and of a sequence of functionals (\ref{f}) of this operator (\ref{ske}).

\textcolor{blue!50!black}{\subsection{Scaling properties of state evolution}}
The generally accepted philosophy of describing evolution by kinetic equations is as follows \cite{Bog},\cite{G58}.
If the initial state is determined by the state of a typical particle of the system, i.e. at the initial moment
there are no correlations between particles (chaos condition), then in a certain scaling approximation
\cite{BPS},\cite{EShY10},\cite{G12},\cite{G13} the evolution of the state of a system of many particles can be
effectively described within the state of a typical particle, i.e. by a one-particle density operator, which is
governed by the corresponding nonlinear kinetic equation.

Further, the scaling asymptotic behavior of the reduced functionals of the state $F(t\mid F_{1}(t))$ in the specific
case of the mean-field limit is considered \cite{GTrmp}.

Let there exists the mean-field limit of the initial one-particle density operator in the following sense
\begin{eqnarray}\label{asic1}
   &&\lim\limits_{\epsilon\rightarrow 0}\big\|\epsilon\,F_{1}^{0}-
      f_{1}^{0}\big\|_{\mathfrak{L}^{1}(\mathcal{H})}=0,
\end{eqnarray}
where $\epsilon>0$ is a scaling parameter.

Since for an arbitrary finite time interval for an asymptotically perturbed first-order cumulant of groups
of operators (\ref{grGs}), i.e. for a strongly continuous group (\ref{grGs}), such equality holds
\begin{eqnarray*}
    &&\lim\limits_{\epsilon\rightarrow 0}\Big\|\mathcal{G}^{\ast}_{s}(t,1,\ldots,s)f_{s}-
        \prod\limits_{j=1}^{s}\mathcal{G}^{\ast}_{1}(t,j)f_{s}\Big\|_{\mathfrak{L}^{1}(\mathcal{H}_{s})}=0.
\end{eqnarray*}
then for the $(1+n)$-th order cumulant of asymptotically perturbed groups of operators (\ref{grGs})
the equality valid
\begin{eqnarray}\label{apgc}
    &&\hskip-12mm\lim\limits_{\epsilon\rightarrow 0}\Big\|\frac{1}{\epsilon^n}\,
       \frac{1}{n!}\mathfrak{A}_{1+n}(t,\{1,\ldots,s\},s+1,\ldots,s+n)f_{s+n}-
\end{eqnarray}
\begin{eqnarray}
    &&\hskip-2mm\int\limits_0^tdt_{1}\ldots\int\limits_0^{t_{n-1}}dt_{n}
        \prod\limits_{j=1}^{s}\mathcal{G}_{1}^{\ast}(t-t_{1},j)
        \sum\limits_{i_{1}=1}^{s}\mathcal{N}^{\ast}_{\mathrm{int}}(i_{1},s+1)
        \prod\limits_{j_1=1}^{s+1}\mathcal{G}_{1}^{\ast}(t_{1}-t_{2},j_1)\ldots\nonumber\\
    &&\hskip-2mm\prod\limits_{j_{n-1}=1}^{s+n-1}\mathcal{G}^{\ast}_{1}(t_{n-1}-t_{n},j_{n-1})
        \sum\limits_{i_{n}=1}^{s+n-1}\mathcal{N}^{\ast}_{\mathrm{int}}(i_{n},s+n)
        \prod\limits_{j_n=1}^{s+n}\mathcal{G}^{\ast}_{1}(t_{n},j_n)f_{s+n}\Big\|_{\mathfrak{L}^{1}(\mathcal{H}_{s+n})}=0,
        \quad n\geq1.\nonumber
\end{eqnarray}
The last equations are a consequence of the validity for the bounded interaction potential of analogues
of the Duhamel equation for cumulants \eqref{cumulant1+n} of the groups of operators (\ref{grGs}).

According to equalities (\ref{apgc}) for a one-particle density operator (\ref{ske}) the following limit
theorem holds.

If condition (\ref{asic1}) is satisfied, then for series (\ref{ske}) the equality is valid
\begin{eqnarray*}
   &&\lim\limits_{\epsilon\rightarrow 0}\big\|\epsilon F_{1}(t)-
     f_{1}(t)\big\|_{\mathfrak{L}^{1}(\mathcal{H})}=0,
\end{eqnarray*}
where for an arbitrary finite time interval the limiting one-particle density operator $f_1(t)$
is determined by convergent series over the norm of space $\mathfrak{L}^{1}(\mathcal{H})$
\begin{eqnarray}\label{viterc}
     &&\hskip-12mm f_{1}(t,1)=\sum\limits_{n=0}^{\infty}\,\int\limits_0^tdt_{1}\ldots
        \int\limits_0^{t_{n-1}}dt_{n}\,\mathrm{Tr}_{2,\ldots,n+1}\mathcal{G}^{\ast}_{1}(t-t_{1},1)
        \mathcal{N}^{\ast}_{\mathrm{int}}(1,2)\prod\limits_{j_1=1}^{2}\mathcal{G}^{\ast}_{1}(t_{1}-t_{2},j_1)\ldots\\
     &&\prod\limits_{i_{n}=1}^{n}\mathcal{G}^{\ast}_{1}(t_{n}-t_{n},i_{n})
         \sum\limits_{k_{n}=1}^{n}\mathcal{N}^{\ast}_{\mathrm{int}}(k_{n},n+1)
         \prod\limits_{j_n=1}^{n+1}\mathcal{G}^{\ast}_{1}(t_{n},j_n)\prod\limits_{i=1}^{n+1}f_1^0(i).\nonumber
\end{eqnarray}
For bounded interaction potentials, the series (\ref{viterc})the series converges in the norm of space
$\mathfrak{L}^{1}(\mathcal{H})$
provided that: $t<t_{0}\equiv(2\,\|\Phi\|_{\mathfrak{L}(\mathcal{H}_{2})}\|f_1^0\|_{\mathfrak{L}^{1}(\mathcal{H})})^{-1}$.

For the initial state $f_{1}^0\in\mathfrak{L}^{1}(\mathcal{H})$ limiting operator \eqref{viterc} satisfies
the Cauchy problem for the quantum Vlasov kinetic equation
\begin{eqnarray}\label{Vlasov1}
     &&\hskip-5mm \frac{\partial}{\partial t}f_{1}(t,1)=\mathcal{N}^{\ast}_{1}(1)f_{1}(t,1)+
       \mathrm{Tr}_{2}\,\mathcal{N}^{\ast}_{\mathrm{int}}(1,2)f_{1}(t,1)f_{1}(t,2),\\ \nonumber\\
   \label{Vlasov2}
     &&\hskip-5mm f_{1}(t)|_{t=0}=f_{1}^0.
\end{eqnarray}

We note that for pure states, i.e. $f_{1}(t)=|\psi(t)\rangle\langle\psi(t)|$ or in terms of the kernel
of this operator $f_{1}(t,q,q')=\psi(t,q)\psi^{\ast}(t,q')$,  the quantum Vlasov kinetic \eqref{Vlasov1}
is reduced to the Hartree equation
\begin{eqnarray*}
    &&i\frac{\partial}{\partial t} \psi(t,q)=-\frac{1}{2}\Delta_{q}\psi(t,q)+
       \int dq'\Phi(q-q')|\psi(q')|^{2}\psi(t,q).
\end{eqnarray*}
In particular, if the kernel of the interaction potential $\Phi(q)=\delta(q)$ is a Dirac measure, then
the Hartree kinetic equation turns into the nonlinear Schr\"{o}dinger equation with cubic nonlinearity
\begin{eqnarray*}
    &&i\frac{\partial}{\partial t}\psi(t,q)=
       -\frac{1}{2}\Delta_{q}\psi(t,q)+ |\psi(t,q)|^{2}\psi(t,q).
\end{eqnarray*}

According to property \eqref{apgc} for cumulants of asymptotically perturbed groups of operators for generating
operators \eqref{skrrc} the equalities true:
\begin{eqnarray*}
    &&\hskip-5mm \lim\limits_{\epsilon\rightarrow 0}\Big\|\big(\mathfrak{V}_{1}(t,\{1,\ldots,s\})-I\big)f_{s}
         \Big\|_{\mathfrak{L}^{1}(\mathcal{H}_{s})}=0,
\end{eqnarray*}
and for $n\geq1$
\begin{eqnarray*}
    &&\hskip-5mm \lim\limits_{\epsilon\rightarrow 0}\Big\|\frac{1}{\epsilon^{n}}\,
        \mathfrak{V}_{1+n}(t,\{1,\ldots,s\},s+1,\ldots,s+n)f_{s+n}
        \Big\|_{\mathfrak{L}^{1}(\mathcal{H}_{s+n})}=0,
\end{eqnarray*}
respectively.

Due to such equalities and convergence in the mean-field limit of the solution of the Cauchy problem
of the generalized quantum kinetic equation \eqref{gkec},\eqref{gkeci} to the solution of the Cauchy
problem of the quantum Vlasov kinetic equation \eqref{Vlasov1},\eqref{Vlasov2} for reduced functionals
of the state (\ref{f}) the following equalities are valid:
\begin{eqnarray*}
    &&\lim\limits_{\epsilon\rightarrow 0}\Big\|\epsilon^{s} F_{s}\big(t,1,\ldots,s \mid F_{1}(t)\big)-
       \prod\limits_{j=1}^{s}f_{1}(t,j)\Big\|_{\mathfrak{L}^{1}(\mathcal{H}_{s})}=0, \quad s\geq2,
\end{eqnarray*}
where the limit operator $f_{1}(t)$ is determined by series (\ref{viterc}), which represents the solution
of the quantum Vlasov equation \eqref{Vlasov1}.

The last statement describes the process of propagation of the initial chaos in the mean-field limit, i.e.,
if at the initial moment there are no correlations in the system, then in the process of evolution correlations
are not created in this approximation.

Note that the traditional approach to the problem of the propagation of initial chaos is based on the construction
of the asymptotic behavior of a solution of the quantum BBGKY hierarchy for reduced density operators within the
perturbation theory \cite{BPS},\cite{G13}.

\textcolor{blue!50!black}{\section{The kinetic evolution of observables}}

In this section, we consider the scaling asymptotic behavior of a solution of the Cauchy problem
\eqref{sdh} for the dual BBGKY hierarchy \eqref{dh},\eqref{dhi} in the case of the mean-field
limit \cite{G11}, or, in other words, we consider the foundations of the description of the kinetic
evolution of quantum systems of many particles within the framework of observables.

Note that one of the advantages of such an approach to the description of kinetic evolution is the
possibility to describe the propagation of initial correlations in scaling limits.

\textcolor{blue!50!black}{\subsection{The hierarchy of kinetic equations for observables}}
Suppose that at the initial moment of time there exists the mean-field limit of reduced observables \eqref{dhi}
in the sense of $\ast$-weak convergence of the space of bounded operators $\mathfrak{L}(\mathcal{H}_s)$
\begin{eqnarray}\label{asumdino}
    &&\mathrm{w^{\ast}-}\lim\limits_{\epsilon\rightarrow 0}\big(\epsilon^{-s}B_{s}^{\epsilon,0}-b_{s}^0\big)=0,
\end{eqnarray}
where $\epsilon>0$ is a scaling parameter.

Then the following limit theorem holds for reduced observables \eqref{sdh}, which are the solution of the dual
BBGKY hierarchy \eqref{dh}.

If the condition \eqref{asumdino} is satisfied, then for an arbitrary finite time interval there is a mean-field
limit for the sequence of reduced observables (\ref{sdh}) in the same sense \cite{G15},\cite{G11}
\begin{eqnarray}\label{asymto}
    &&\mathrm{w^{\ast}-}\lim\limits_{\epsilon\rightarrow 0}\big(\epsilon^{-s}B_{s}(t)-b_{s}(t)\big)=0,
\end{eqnarray}
where the reduced observables $b_{s}(t),\,s\geq1$, are determined by the following expansions:
\begin{eqnarray}\label{Iterd}
   &&\hskip-12mm  b_{s}(t,1,\ldots,s)=\sum\limits_{n=0}^{s-1}\,\int\limits_0^tdt_{1}\ldots\int\limits_0^{t_{n-1}}dt_{n}
      \,\mathcal{G}_s^{0}(t-t_{1})\sum\limits_{i_{1}\neq j_{1}=1}^{s}
      \mathcal{N}_{\mathrm{int}}(i_{1},j_{1})\mathcal{G}_{s-1}^{0}(t_{1}-t_{2})\ldots\\
   &&\hskip-12mm \mathcal{G}_{s-n+1}^{0}(t_{n-1}-t_{n})
      \sum\limits^{s}_{\mbox{\scriptsize $\begin{array}{c}i_{n}\neq j_{n}=1,\\
      i_{n},j_{n}\neq(j_{1},\ldots,j_{n-1})\end{array}$}}\mathcal{N}_{\mathrm{int}}(i_{n},j_{n}) \mathcal{G}_{s-n}^{0}(t_{n})b_{s-n}^0((1,\ldots,s)\setminus({j_{1}},\ldots,{j_{n}})),\nonumber\\
   &&\hskip-12mm s\geq1,\nonumber
\end{eqnarray}
and for the group of operators of non-interacting particles, the notation was used
\begin{eqnarray*}
   && \mathcal{G}_{s-n+1}^{0}(t_{n-1}-t_{n})\equiv
      \prod\limits_{j\in (1,\ldots,s)\setminus (j_1,\ldots,j_{n-1})}\mathcal{G}_{1}(t_{n-1}-t_{n},j).
\end{eqnarray*}

For a certain type of observables, the structure of expansion (\ref{Iterd}) takes a particular form,
for example, in the case of the $k$-ary type of observables we have:
\begin{eqnarray}\label{kIterd}
   &&\hskip-12mm  b_{s}^{(k)}(t,1,\ldots,s)=\int\limits_0^tdt_{1}\ldots\int\limits_0^{t_{s-k-1}}dt_{s-k}
      \,\mathcal{G}_s^{0}(t-t_{1})\sum\limits_{i_{1}\neq j_{1}=1}^{s}\mathcal{N}_{\mathrm{int}}(i_{1},j_{1})
      \mathcal{G}_{s-1}^{0}(t_{1}-t_{2})\ldots\\
   && \mathcal{G}_{s-n+1}^{0}(t_{s-k-1}-t_{s-k})\sum\limits^{s}_{\mbox{\scriptsize $\begin{array}{c}i_{s-k}\neq j_{s-k}=1,\\
      i_{s-k},j_{s-k}\neq (j_{1},\ldots,j_{s-k-1})\end{array}$}}\mathcal{N}_{\mathrm{int}}(i_{s-k},j_{s-k})
      \mathcal{G}_{s-n}^{0}(t_{s-k})\times\nonumber\\
   &&  b_{k}^0((1,\ldots,s)\setminus({j_{1}},\ldots,{j_{s-k}})),\quad  1\leq s\leq k.\nonumber
\end{eqnarray}

If $b^0\in \mathfrak{L}(\mathcal{F}_{\mathcal{H}})$, then the sequence $b(t)=(b_0,b_1(t),\ldots,b_{s}(t),\ldots)$
of limit observables (\ref{Iterd}) satisfies the Cauchy problem of the dual Vlasov hierarchy:
\begin{eqnarray}\label{vdh}
   &&\hskip-12mm \frac{\partial}{\partial t}b_{s}(t,1,\ldots,s)=\sum\limits_{j=1}^{s}\mathcal{N}(j)\,b_{s}(t,1,\ldots,s)+
     \sum_{j_1\neq j_{2}=1}^s\mathcal{N}_{\mathrm{int}}(j_1,j_{2})\,b_{s-1}(t,(1,\ldots,s)\setminus(j_1)),\\
\nonumber\\
\label{vdhi}
   &&\hskip-12mm b_{s}(t,1,\ldots,s)\mid_{t=0}=b_{s}^0(1,\ldots,s),\quad s\geq1.
\end{eqnarray}

We give some examples of the dual Vlasov hierarchy \eqref{vdh} in terms of kernels of the operators for the reduced observables:
\begin{eqnarray*}
    &&\hskip-8mm i\,\frac{\partial}{\partial t}b_{1}(t,q_1;q'_1)=
       -\frac{1}{2}(-\Delta_{q_1}+\Delta_{q'_1})b_{1}(t,q_1;q'_1),\\
    &&\hskip-8mm i\,\frac{\partial}{\partial t}b_{2}(t,q_1,q_2;q'_1,q'_2)=
       \big(-\frac{1}{2}\sum\limits_{i=1}^2(-\Delta_{q_i}+\Delta_{q'_i})+\\
    &&\hskip+10mm (\Phi(q'_1-q'_2)-\Phi(q_1-q_2))\big )b_{2}(t,q_1,q_2;q'_1,q'_2))+\\
    &&\hskip+10mm \big(\Phi(q'_1-q'_2)-\Phi(q_1-q_2)\big)\big(b_{1}(t,q_1;q'_1)+b_{1}(t,q_2;q'_2)\big).
\end{eqnarray*}
We note that the sequence of evolution equations (\ref{vdh}) has the structure of recursive equations.

Thus, in the mean-field limit, the collective behavior (kinetic evolution) of quantum systems
of many particles is described in terms of the sequence of limiting reduced observables (\ref{Iterd}),
in particular, by sequence (\ref{kIterd}) whose evolution is governed by the Cauchy problem of the
dual Vlasov hierarchy (\ref{vdh}),(\ref{vdhi}).

\textcolor{blue!50!black}{\subsection{The propagation of the initial chaos}}
Let us consider the relationship of collective behavior within the mean-field approximation of a quantum
system of many particles, which is described by the dual Vlasov hierarchy \eqref{vdh} for the limiting
reduced observables, and by the Vlasov kinetic equation  \eqref{Vlasov1} for the state of a typical
particle of a system.

Let the initial state satisfy a chaos condition, i.e. at the initial instant  there are no correlations
between particles (the statistically independent particles), namely, in the case of the Maxwell--Boltzmann
statistics by such a sequence of limiting reduced density operators
\begin{eqnarray}\label{h}
      && f^{(c)}\equiv\big(I,f_1^{0}(1),\ldots,\prod_{i=1}^s f_1^{0}(i),\ldots\big).
\end{eqnarray}
As mentioned above, this assumption regarding the initial state is characteristic of the kinetic description
of the gas, because in this case the state is completely determined using a one-particle density operator.

For mean value functional \eqref{avmar} of additive type limit observables (\ref{kIterd}) and  initial state
\eqref{h} on a finite time interval, the equality holds
\begin{eqnarray*}\label{avmar-2}
  &&\hskip-8mm\big(b^{(1)}(t),f^{(c)}\big)=\sum\limits_{s=0}^{\infty}\frac{1}{s!}\,
         \mathrm{Tr}_{1,\ldots,s}\,b_{s}^{(1)}(t,1,\ldots,s)\prod\limits_{i=1}^{s}f_{1}^0(i)=\\
  &&\hskip+7mm \mathrm{Tr}_{1}\,b_{1}^{0}(1)f_{1}(t,1),\nonumber
\end{eqnarray*}
where a one-particle density operator is represented by such a series
\begin{eqnarray}\label{viter}
     &&\hskip-12mm f_{1}(t,1)=\sum\limits_{n=0}^{\infty}
        \int\limits_0^tdt_{1}\ldots\int\limits_0^{t_{n-1}}dt_{n}\,
        \mathrm{Tr}_{2,\ldots,n+1}\prod\limits_{i_1=1}^{1}\mathcal{G}^{\ast}_{1}(t-t_{1},i_1)
        \mathcal{N}^{\ast}_{\mathrm{int}}(1,2)\prod\limits_{j_1=1}^{2}
        \mathcal{G}^{\ast}_{1}(t_{1}-t_{2},j_1)\ldots\\
     &&\prod\limits_{i_{n}=1}^{n}\mathcal{G}^{\ast}_{1}(t_{n}-t_{n},i_{n})
        \sum\limits_{k_{n}=1}^{n}\mathcal{N}^{\ast}_{\mathrm{int}}(k_{n},n+1)
        \prod\limits_{j_n=1}^{n+1}\mathcal{G}^{\ast}_{1}(t_{n},j_n)\prod\limits_{i=1}^{n+1}f_1^0(i).\nonumber
\end{eqnarray}
The one-particle density operator \eqref{viter} satisfies the Cauchy problem of the quantum Vlasov
kinetic equation \eqref{Vlasov1},\eqref{Vlasov2}.

Thus, the hierarchy of evolutionary equations (\ref{vdh}) for limit reduced observables of additive type
and initial state (\ref{h}) describes the evolution of quantum many-particle systems in an equivalent
way compared to the Vlasov equation \eqref{Vlasov1}.

Accordingly, for the mean value functional of the limit observables of nonadditive type (\ref{kIterd})
and the initial state \eqref{h} on a finite time interval, equations valid:
\begin{eqnarray}\label{pchaos}
    &&\hskip-7mm\big(b^{(k)}(t),f^{(c)}\big)=\sum\limits_{s=0}^{\infty}\frac{1}{s!}\,
         \mathrm{Tr}_{1,\ldots,s}\,b_{s}^{(k)}(t,1,\ldots,s)\prod\limits_{i=1}^{s}f_1^0(i)=\\
    &&\frac{1}{k!}\mathrm{Tr}_{1,\ldots,k}\,b_{k}^{0}(1,\ldots,k)
       \prod\limits_{i=1}^{k}f_{1}(t,i),\quad k\geq2,\nonumber
\end{eqnarray}
where the one-particle density operator is represented by series (\ref{viter}), i.e. satisfies
the Cauchy problem for the quantum Vlasov kinetic equation \eqref{Vlasov1},\eqref{Vlasov2}.

Equality \eqref{pchaos} describes the process of the propagation of initial chaos (\ref{h}) in
the mean-field limit by means of solution (\ref{viter}) of  the quantum Vlasov kinetic eqation
\eqref{Vlasov1}, namely, for an arbitrary finite time interval, the state is represented by such
a sequence of limit reduced operators:
\begin{eqnarray}\label{ihaos}
     &&f_k(t,1,\ldots,k)=\prod\limits_{i=1}^{k}f_{1}(t,i),\quad k\geq2,
\end{eqnarray}
that is, if at the initial moment of time in the system of particles there are no correlations, then
in this approximation in the process of evolution correlations of particle states are not create.

\textcolor{blue!50!black}{\subsection{Kinetic equations with initial correlations}}
We note that the above approach to the derivation of kinetic equations allows us to formulate kinetic
equations in the case of more general initial states, which describe not only the gases of quantum
particles (\ref{h}) but also systems of many particles in condensed states.

Further, we consider the initial states of quantum systems of many particles, which are determined by
a one-particle density operator and correlation operators (Maxwell - Boltzmann statistics) \cite{G15}
\begin{eqnarray}\label{insc}
   &&\hskip-12mm f^{(cc)}=\big(I,f_1^{0}(1),g_{2}^0(1,2)\prod_{i=1}^{2}f_1^{0}(i),
        \ldots,g_{n}^0(1,\ldots,n)\prod_{i=1}^{n}f_1^{0}(i),\ldots\big),
\end{eqnarray}
where the correlations of the initial states of the particles are determined by the operators
$g_{n}^0(1,\ldots,n)\equiv g_{n}^0\in\mathfrak{L}^{1}_0(\mathcal{H}_n),\,n\geq2$. We emphasize
that this assumption (\ref{insc}) with respect to the initial state is typical for the kinetic
description of systems of many particles in condensed states, which are characterized by correlations,
for example, such as in particle flows \cite{BogLect},\cite{SR04}.

Then, using the method of derivation of kinetic equations based on the hierarchy of kinetic equations
for the observables, for an arbitrary finite time interval we establish that the state is described by
the sequence $f(t)=\big(I,f_{1}(t),\ldots,f_{n}(t,1,\ldots,n),\ldots\big)$ of limiting reduced density
operators where a one-particle density operator is represented by such a series
\begin{eqnarray}\label{vitercc}
   &&\hskip-12mm f_{1}(t,1)=\sum\limits_{n=0}^{\infty}
        \int\limits_0^tdt_{1}\ldots\int\limits_0^{t_{n-1}}dt_{n}\,
        \mathrm{Tr}_{2,\ldots,n+1}\prod\limits_{i_1=1}^{1}\mathcal{G}^{\ast}_{1}(t-t_{1},i_1)
        \mathcal{N}^{\ast}_{\mathrm{int}}(1,2)\prod\limits_{j_1=1}^{2}
        \mathcal{G}^{\ast}_{1}(t_{1}-t_{2},j_1)\ldots\\
     &&\prod\limits_{i_{n}=1}^{n}\mathcal{G}^{\ast}_{1}(t_{n}-t_{n},i_{n})
        \sum\limits_{k_{n}=1}^{n}\mathcal{N}^{\ast}_{\mathrm{int}}(k_{n},n+1)
        \prod\limits_{j_n=1}^{n+1}\mathcal{G}^{\ast}_{1}(t_{n},j_n)g_{n+1}^0(1,\ldots,n+1)
        \prod\limits_{i=1}^{n+1}f_1^0(i),\nonumber
\end{eqnarray}
and the limiting density operators $f_k(t,1,\ldots,k),\,k\geq2,$ are determined by the following
expressions \cite{GTsm}:
\begin{eqnarray}\label{dchaos}
     &&\hskip-12mm f_k(t,1,\ldots,k)=\prod_{i_1=1}^{k}\mathcal{G}^{\ast}_{1}(t,i_1)
       g_{k}^0(1,\ldots,k)\prod_{i_2=1}^{k}(\mathcal{G}_{1}^{\ast})^{-1}(t,i_2)
       \prod\limits_{j=1}^{k}f_{1}(t,j),\quad k\geq2.
\end{eqnarray}

Indeed, for the mean value functionals of the limiting observables (\ref{kIterd}) in the case of
the initial state \eqref{insc} for a finite time interval, the equalities are valid:
\begin{eqnarray*}
    &&\hskip-7mm \big(b^{(k)}(t),f^{(cc)}\big)=
       \frac{1}{k!}\mathrm{Tr}_{1,\ldots,k}\,b_{k}^{0}(1,\ldots,k)f_{k}(t,1,\ldots,k),\quad k\geq1,
\end{eqnarray*}
where the sequence $f(t)$ is determined by expansions \eqref{vitercc} and \eqref{dchaos}.

In \cite{GTsm}, a similar result was obtained using a generalized quantum kinetic equation with
initial correlations.

In the case of initial states specified by a sequence of limiting correlation operators (\ref{insc}),
the process of propagation of initial correlations is described by the following sequence of correlation
operators (\ref{gBigfromDFB}):
\begin{eqnarray*}
   &&\hskip-12mm g_n(t,1,\ldots,n)=\prod_{i_1=1}^{n}\mathcal{G}_{1}^{\ast}(t,i_1)
      \sum\limits_{\mbox{\scriptsize$\begin{array}{c}\mathrm{P}:(1,\ldots,n)=\bigcup_{i}X_{i}\end{array}$}}
      (-1)^{|\mathrm{P}|-1}(|\mathrm{P}|-1)!\,\prod_{X_i\subset\mathrm{P}}g_{|X_i|}^0(X_{i})\times\\
   && \prod_{i_2=1}^{n}(\mathcal{G}_{1}^{\ast})^{-1}(t,i_2)
      \prod\limits_{j=1}^{n}f_{1}(t,j), \quad n\geq2.
\end{eqnarray*}

Limiting one-particle density operator (\ref{vitercc}) is governed by the quantum Vlasov kinetic
equation with initial correlations \cite{GTsm}:
\begin{eqnarray}\label{Vls}
  &&\hskip-12mm \frac{\partial}{\partial t}f_{1}(t,1)=\mathcal{N}^{\ast}(1)f_{1}(t,1)+\\
  && \mathrm{Tr}_{2}\,\mathcal{N}^{\ast}_{\mathrm{int}}(1,2)
     \prod_{i_1=1}^{2}\mathcal{G}^{\ast}_{1}(t,i_1)g_{2}^0(1,2)
     \prod_{i_2=1}^{2}(\mathcal{G}^{\ast}_{1})^{-1}(t,i_2)f_{1}(t,1)f_{1}(t,2),\nonumber\\
\label{Vlasov2c}
  &&\hskip-12mm f_{1}(t)|_{t=0}=f_{1}^0,
\end{eqnarray}
where the notations \eqref{infOper1} is used and the inverse group of operators to group
(\ref{grG}) is denoted by $(\mathcal{G}^{\ast}_{1})^{-1}(t)$.

We note that the kinetic equation (\ref{Vls}) is a non-Markov kinetic equation. For pure states,
equation (\ref{Vls}) reduces to the Hartree kinetic equation with initial correlations. For the
initial states of the system of statistically independent particles, the kinetic equation (\ref{Vls})
coincides with the quantum Vlasov equation \eqref{Vlasov1}, and reduced density operators (\ref{dchaos})
describe the process of propagation of the initial chaos (\ref{ihaos}).

We remark also that in the case of arbitrary initial states in a  mean-field limit, a sequence of reduced
density operators, as well as sequence (\ref{vitercc}), (\ref{dchaos}), is the solution of the Cauchy
problem of the Vlasov hierarchy:
\begin{eqnarray*}
     &&\hskip-12mm \frac{\partial}{\partial t}f_{s}(t,1,\ldots,s)=
        \sum\limits_{i=1}^{s}\mathcal{N}^{\ast}(i)f_{s}(t,1,\ldots,s)+
        \sum\limits_{i=1}^{s}\mathrm{Tr}_{s+1}\mathcal{N}^{\ast}_{\mathrm{int}}(i,s+1)f_{s+1}(t,1,\ldots,s+1),\\
        \\
     &&\hskip-12mm f_{s}(t)_{|t=0}=f_{s}^{0},\quad s\geq1.
\end{eqnarray*}

Thus, in the case of initial states given by a one-particle density operator and correlation operators
(\ref{insc}), the dual Vlasov hierarchy (\ref{vdh}) for the additive type reduced observables describes
the evolution of quantum systems of many particles, just as non-Markov quantum Vlasov kinetic equation
with initial correlations (\ref{Vls}). In the case of reduced nonadditive-type observables, the dual
Vlasov hierarchy (\ref{vdh}) describes in an equivalent way in the sense of equality (\ref{dchaos})
the process of propagation of initial correlations in terms of reduced density operators (\ref{dchaos}).
In other words, the alternative method of describing the evolution of states of quantum systems of many
particles in the mean-field approximation is based on the non-Markovian Vlasov kinetic equation with
initial correlations (\ref{Vls}).


\textcolor{blue!50!black}{\section{Outlook}}

The possible approaches to the description of the evolution of quantum systems of many particles
are considered above, namely, in terms of the observables ones, the evolution of which is described
by the dual BBGKY hierarchy (\ref{dh}), or in terms of the state whose evolution is described by
the BBGKY hierarchy (\ref{BBGKY}). For systems of a finite number of particles, these hierarchies
of evolution equations are equivalent to the Heisenberg equation \eqref{H-N1} and the von Neumann
equation \eqref{vonNeumannEqn}, respectively. In particular, possible methods of describing the
evolution of the state by means of the hierarchy of fundamental equations \eqref{vNh}, which
describes the evolution of state correlations, are considered.

It was established that the concept of cumulant (\ref{cumcp}) of groups of operators forms the
basis of expansions for solutions of fundamental evolution equations, which describe the evolution
of quantum systems of many particles, namely: in the case of the groups of operators \eqref{grG} for
the dual BBGKY hierarchy for reduced observables; in the case of groups of operators (\ref{grGs}) for
the von Neumann hierarchy (\ref{vNh}) for correlation operators, for the BBGKY hierarchy (\ref{BBGKY})
for reduced density operators and for the BBGKY hierarchy of nonlinear equations (\ref{gBigfromDFBa})
for reduced correlation operators, respectively, as well as for the basis of the kinetic description
of infinitely many particles systems \eqref{f},(\ref{gkec}).

We emphasize that the structure of expansions for correlation operators (\ref{rozvNF-Nclusters}), in
which the generating operators are of the appropriate order cumulants (\ref{cumulantP}) of the groups
of operators (\ref{grGs}), induces the cumulant structure of expansions in a series for reduced density
operators (\ref{RozvBBGKY}), reduced correlation operators (\ref{GClusters}) and reduced functionals of
the state (\ref{f}). Thus, the dynamics of systems of infinitely many particles are generated by the
dynamics of correlations.

The article also was considered two new approaches to describing the kinetic evolution of quantum systems
of many particles. One of them consists in the description of the kinetic evolution of the quantum system
of particles by means of the reduced observables in mean-field scaling limit \eqref{vdh}. Another approach
is based on the non-Markovian generalization of quantum kinetic equations \eqref{gkec}.

One of the advantages of the considered approaches is connected with the possibility of construction
of kinetic equations in scaling approximations taking into account correlations of particles at the
initial instant which characterize the condensed states of systems of many particles. Another advantage
is related to the problem of the rigorous derivation of kinetic equations of the non-Markovian type based
on the dynamics of quantum many-particle systems, which allow us to describe the effects of memory in
nanostructures.

We remark that the approach to the derivation of the quantum Vlasov kinetic equation from the dynamics
of many-particle systems, which is based on the generalized quantum kinetic equation \eqref{gkec}, also
allows the construction of higher-order corrections to the approximation of mean-field for quantum kinetic
equations.

The above results can be extended to systems of many bosons or fermions \cite{GP},\cite{GTs10}.


\addcontentsline{toc}{section}{References}
\renewcommand{\refname}{References}

\end{document}